\documentclass[aps,prd,twocolumn,showpacs,amsmath,amssymb]{revtex4}

\usepackage{graphicx}
\usepackage{dcolumn}
\usepackage{bm}

\begin{document}

\title{Born-Infeld  black hole in the isolated horizon framework}
\author{Nora Bret\'on}
\email{nora@fis.cinvestav.mx}
\affiliation{Departamento de F\'{\i}sica,
Cinvestav--IPN, Apartado Postal 14--740, C.P. 07000, M\'exico,
D.F., MEXICO}

\begin{abstract}
In this work we probe the Born-Infeld (BI) black hole in the
isolated horizon framework. It turns out that the BI black
hole is consistent with the heuristic model for colored
black holes proposed by Ashtekar {\it et al} 
[(2001){\it Class.Quant.Grav.} {\bf18} 919-940].
The model points to the unstability of the BI black hole.
\end{abstract} 

\maketitle


\section{Introduction}

The Reissner-Nordstr\"om (RN) solution (characterized by its charge
and mass) turns out to be the final fate of a charged star, having as
uncharged limit the Schwarzschild black hole, then it is of interest to
investigate in more detail its nonlinear electromagnetic generalization,
in particular, the Born-Infeld field \cite{Born}.
 
The Einstein-Born-Infeld (EBI) generalization of the RN black hole was
obtained by Garc\'{\i}a-Salazar-Pleba\~nski (GSP) in 1984 \cite{GSP}.  Two
years after, Demianski \cite{Geon} presented the static spherically symmetric
solution that nowadays is known as the EBIon, it is the most known solution of
the EBI equations; in the spirit of the concept of geon introduced by Wheeler
in the sixties, this means electromagnetic radiation held together by its
self- gravitational attraction. These two solutions, GSP and Demianski's, are
actually the same, differing only by a constant.
 
Remarkable properties of the EBI black hole arise in the context of the
isolated horizon formalism, recently put forward by Ashtekar and co-workers
in a series of papers in Phys. Rev. D and Class. and Q. Grav.
\cite{Ashtekar}. In this approach it is pointed out the unsatisfactory
(uncomplete) description of a black hole given by concepts such as ADM mass
and event horizon, for instance, specially if one is dealing with hairy
black holes. To remedy this uncompleteness, Ashtekar et al have proposed
alternatively the isolated horizon formalism, that furnish a more complete
description of what happens in the neighborhood of the horizon of a hairy
black hole. Moreover, they conjecture about the relationship between the
colored black holes and their solitonic analogs \cite{Sudarsky}: the ADM
mass contains two contributions, one attributed to the black hole horizon
and the other to the outside hair, captured by the solitonic residue. In
the present communication we show that the EBI colored black hole and the
corresponding solitonic solution have most of the properties of the model
proposed in \cite{Sudarsky}.

The EBI black hole corresponds to the solution for the field equations
arising from the Einstein-Born-Infeld action

\begin{equation}
S = \int{d^4x \sqrt{-g} \{ R (16\pi)^{-1}+L \} } ,
\end{equation} 
where $R$ denotes the scalar curvature, $g:= {\rm det} \vert g_{\mu \nu}
\vert$ and $L$, the electromagnetic part, is assumed to depend in nonlinear
way on the invariants of $P_{\mu \nu}$, the nonlinear generalization of the
electromagnetic field, $F_{\mu \nu}$,

\begin{equation}
L=-{1 \over 2} P^{\mu \nu} F_{\mu \nu} +K(P,Q),
\label{Lagr}
\end{equation}
where $P$ and $Q$ are the invariants of $P_{\mu \nu}$, 
$K(P, Q)$  is the so called structural function which
for the Born-Infeld nonlinear electrodynamics is given by
\begin{equation}
K=b^2 \large(1-\sqrt{1-{2P}/{b^2}+ {Q^2}/{b^4}} \large),
\label{BIK}
\end{equation}
where $b$ is the maximum field strenght and the relevant parameter of the
BI theory.
 
The EBI solution for a static spherically symmetric (SSS) spacetime is
given by:

\begin{eqnarray}
ds^2&=& - \psi dt^2+\psi^{-1}dr^2+r^2(d \theta^2 + \sin^2{\theta}d \phi^2), \\
\psi&=& 1-\frac{2m}{r} + \frac{2}{3}b^2 r^2 (1- \sqrt{1+
\frac{a^4}{r^4}})+
\frac{4q^2}{3r}g(r), \\
g'(r)&=&- (r^4+a^4)^{- \frac{1}{2}},
\label{BImetrfunc}
\end{eqnarray}
where $g'(r)=\frac{dg(r)}{dr}$, $m$ is the mass parameter, $q$ is the electric
charge (both in lenght units), $a^4=q^2/b^2$ and $b$ is the Born-Infeld
parameter given in units of $[\rm{lenght}]^{-1}$. The nonvanishing components
of the electromagnetic field are

\begin{equation}
F_{rt}= q (r^4+ a^4)^{- \frac{1}{2}},
\quad P_{rt}= \frac{q}{r^2}. 
\label{FrtBI}
\end{equation}

The solution given by Garc\'{\i}a-Salazar-Pleba\~nski \cite{GSP} corresponds
to

\begin{equation}
g(r)=
\int^{\infty}_{r}{\frac{ds}{\sqrt{s^4+a^4}}}=\frac{1}{2a}
F(\arccos{\{ \frac{r^2- a^2}{r^2+a^2}} \},\frac{1}{\sqrt{2}}),
\label{gPleb}
\end{equation}
while the one given by Demianski \cite{Geon} corresponds to

\begin{equation}
g(r)= - \int^{r}_{0}{\frac{ds}{\sqrt{s^4+a^4}}}=- \frac{1}{2a}
F(\arccos{\{ \frac{a^2-r^2}{a^2+r^2}} \},\frac{1}{\sqrt{2}}).
\label{gDem}
\end{equation}
  
Choosing $g(r)$ as in Eq. (\ref{gPleb}) or Eq. (\ref{gDem})  has as a
consequence a different behavior of the solution at the origin.  The metric
function $\psi$ with $g(r)$ given in Eq. (\ref{gPleb}) (GSP solution) diverges
at $r \to 0$ (even when $m=0$), it corresponds to the black hole solution. The
other one, meaning $\psi$ with $g(r)$ given in Eq. (\ref{gDem}) (Demianski),
is the so called EBIon, a particlelike solution that is finite at the origin
(for $m=0$). The integrals of Eqs. (\ref{gPleb}) and (\ref{gDem}) are related
by

\begin{equation}
\int^{\infty}_{r}{\frac{ds}{\sqrt{s^4+a^4}}}= - \int^{r}_{0}
{\frac{ds}{\sqrt{s^4+a^4}}}+\rm{Const},
\label{relint}
\end{equation}

The constant can be fixed by imposing that in the limit $b \to \infty$ we
recover the RN solution, then,

\begin{equation}
{\rm{Const}}=\int^{\infty}_{r}{\frac{ds}{\sqrt{s^4+a^4}}}+\int^{r}_{0}
{\frac{ds}{\sqrt{s^4+a^4}}}=\frac{1}{a}{\rm K}(\frac{1}{2})
\end{equation}
where ${\rm K}(\frac{1}{2})$ is the complete elliptic integral of the
first kind. In the limit of large distances, $r \to \infty$,
asymptotically the solution corresponds to the RN solution. Also when the
BI parameter goes to infinity, $b \to \infty$, we recover the linear
electromagnetic (Einstein-Maxwell) RN solution. In the uncharged limit,
$b=0$ (or $q=0$), it is recovered the Schwarzschild black hole.

\subsection{Hairy EBI Black hole}

When the black hole is not completely determined by global charges defined
at spatial infinity such as ADM mass, angular momentum or electric charge,
but rather it possesses short range charges (hair) that vanish at
infinity, then it is a hairy black hole. This is the case for the EBI
black hole, since the location and size of the horizon depends on the
parameter $bq$ (the corresponding metric function was studied in
\cite{Breton}), however, at infinity it is undistinguisable from a RN
black hole characterized only by its charge $q$ and mass $m$.
 
In a series of papers Ashtekar {\it et al} have proposed a more complete
description to characterize a hairy black hole, based on quantities defined
at the horizon. This formalism is intended to deal with situations more
general than SSS hairy black holes and it involves the canonical formalism
of gravity.  Furthermore, they proposed a formula relating the horizon mass
and the ADM mass of the colored black hole solution with the ADM mass of
the soliton solution of the corresponding theory,

\begin{equation}
M^{(n)}_{sol}=M^{(n)}_{ADM} -M^{(n)}_{hor},
\label{massrel}
\end{equation}
 
\noindent where the superscript $n$ indicates the colored version of the
hole; in the papers of Ashtekar {\it et al} this $n$ refers to the
Yang-Mills hair, labeled by this parameter, corresponding to $n=0$ the
Schwarszchild limit (absence of YM charge). This relation has been proved
numerically to work for the Einstein-Yang-Mills (EYM) black hole. For the
case studied here the $n$ version shall correspond to the distinct hairy
black holes labeled by distinct (continuous) BI parameter, $b$; to the
limit $b \to \infty$ corresponds the linearly charged case, that is
precisely the RN black hole; in the limit $b \to 0$ (or $q=0$) we arrive to
the Schwarszchild black hole (our $n=0$ also).
 
It turns out that the EBI black hole and the corresponding EBIon solution
fulfill the relation between the masses as well as most of the properties
of the model for the colored black hole, as will be shown in what follows.
The EBI black hole case has the additional advantage of being an exact
solution.
  
First of all let's check the relation between the masses Eq.
(\ref{massrel}).  The horizon and ADM masses as functions of the horizon
radius $r_h$ for the EBI solution are given, respectively, by

\begin{widetext}
\begin{equation}
M^{(b)}_{hor}(r_h)= \frac{r_h}{2}+\frac{b^2
r_{h}}{3}(r_{h}^2-\sqrt{r_{h}^4+a^4})-\frac{2q^2}{3}
\int^{r_h}_{0}{\frac{ds}{\sqrt{a^4+s^4}}}, 
\label{Mhor}
\end{equation}

\begin{equation}
M^{(b)}_{ADM}(r_h)= \frac{r_h}{2}+\frac{b^2
r_h}{3}(r_{h}^2-\sqrt{r_{h}^4+a^4})+\frac{2q^2}{3}
\int^{\infty}_{r_h}{\frac{ds}{\sqrt{a^4+s^4}}},
\label{Madm}
\end{equation}
\end{widetext}
  
\begin{figure}
\includegraphics[width=8.3cm,height=6cm]{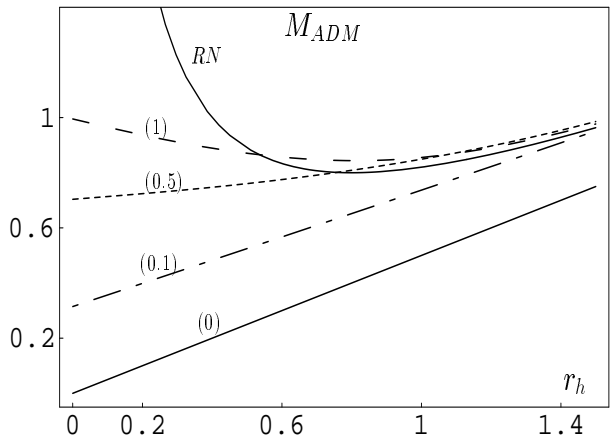}
\caption{\label{biihfig1}
It is shown the ADM mass as function of the horizon radius $r_h$,
$M_{ADM}^{(b)}(r_h)$ for different values of the BI parameter $b$ (in
parenthesis) compared with the corresponding to Reisner-Nordsr\"om,
$M_{ADM}^{RN}(r_h)$, and the bare black hole mass $M_{ADM}^{(0)}(r_h)$
(Schwarszchild).}
\end{figure}

In Fig.\ref{biihfig1} it is shown $M_{ADM}^{(b)}(r_h)$ for different values
of the BI parameter $b$ compared with $M_{ADM}^{RN}(r_h)$ of
Reissner-Nordstr\"om and the bare black hole (Schwarszchild).

The mass of the soliton can be obtained by letting $r \to 0$ in the ADM
mass, Eq. (\ref{Madm}), we obtain $M_{sol}^{(b)}=\frac{2q \sqrt{qb}}{3}{\rm
K}(\frac{1}{2})$. From these expressions one can trivially check that they
satisfy Eq. (\ref{massrel}). Moreover, in the heuristic model for the
colored black hole there are predictions that the EBI solution satisfies.
We enumerate them as stated in \cite{Sudarsky}:

\begin{figure}
\includegraphics[width=8.3cm,height=6cm]{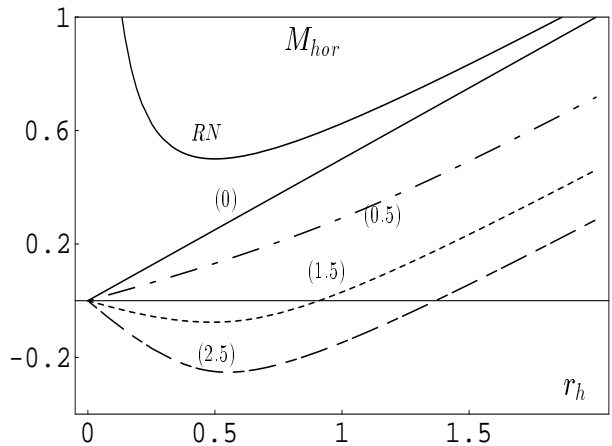}
\caption{\label{biihfig2} 
The inequality satisfied by the horizon masses is shown in this figure for
$b=0.5, b=1.5$ and $b=2.5$, $M_{hor}^{RN}(r_h) > M_{hor}^{(0)}(r_h) >
M_{hor}^{(b)}(r_h)$. }
\end{figure}

(i) 
The bare black hole horizon mass, $M^{(0)}_{hor}(r_h)$, is greater than the
`dressed' mass, $M^{(n)}_{hor}(r_h)$, for all $n$ and all values of the
horizon radius, $r_h$. The corresponding for the EBI case is
$M^{(b)}_{hor}(r_h) < M^{(0)}_{hor}(r_h)$, which amounts to

\begin{equation}
\frac{r_h}{2} > \frac{r_h}{2}- [\frac{b^2r^3}{3}(\sqrt{1+\frac{a^4}{r^4}}-1)+
\frac{2q^2}{3} \int^{r_h}_{0}{\frac{ds}{\sqrt{a^4+s^4}}}],
\end{equation}
 
\noindent since the term in square brackets is positive, the unequality
holds for all $b$, for all $r_h$. Moreover, the horizon masses satisfy the
inequality $M_{hor}^{RN}(r_h) > M_{hor}^{(0)}(r_h) > M_{hor}^{(b)}(r_h)$.
This is shown in Fig.{\ref{biihfig2}}.

\begin{figure}
\includegraphics[width=8.3cm,height=6cm]{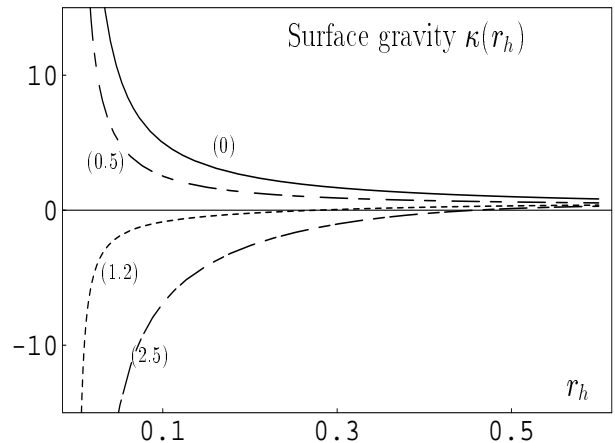}
\caption{\label{biihfig3}   
It is displayed the surface gravity for the BI black hole, $\kappa_{(b)}
(r_h)$ compared with the Schwarszchild surface gravity
$\kappa_{(0)}(r_h)=1/2r_h$. }
\end{figure}

(ii) For all $b$ and all $r_h$, the surface gravity of the coloured black hole
is less that the one for the bare black hole, i.e. $\kappa_{(b)} (r_h) <
\kappa_{(0)}(r_h)$,

\begin{equation} 
\frac{1}{2r_h} \{1+2b^2(r_{h}^2- \sqrt{r_{h}^4+a^4}) \} < \frac{1}{2r_h},
\label{kappa}
\end{equation}
the unequality reduces to 

\begin{equation}
r_{h}^2(1- \sqrt{1+ \frac{a^4}{r^4}}) <0,
\end{equation}
that in fact is satisfied since $\sqrt{1+ \frac{a^4}{r^4}} >1$ for $q, b \neq
0$. It is shown in Fig. \ref{biihfig3} for some values of $b$.

(iii)  From Figs.  {\ref{biihfig2}} and {\ref{biihfig3}} it is manifest
that both $M^{(b)}_{hor}(r_h)$ and $\kappa_{(b)}(r_h)$, for fixed $r_h$,
are monotonically decreasing functions of $b$. In other words, as $ b \to
0$, the coloured black hole tends to the bared one.

\begin{figure}
\includegraphics[width=8.3cm,height=6cm]{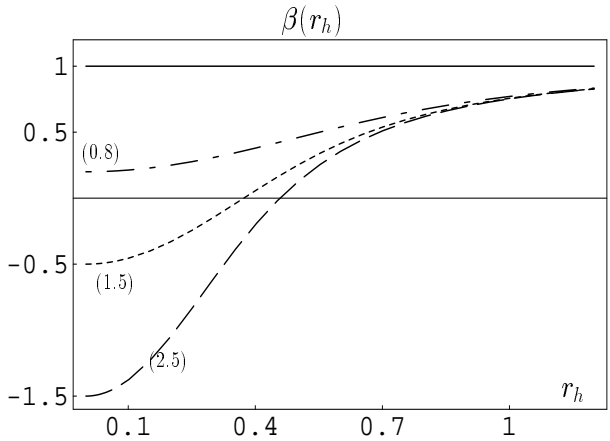}
\caption{\label{biihfig4}
The function $\beta_{(b)}(r_h) = 2r_h \kappa_{(b)}(r_h) < 1$ 
is displayed. For the Schwarzschild black hole, $\beta_0(r_h)=1$.
}
\end{figure}
 
(iv) For fixed $b$, the function $\beta_{(b)}(r_h) = 2r_h \kappa_{(b)}(r_h)
< 1$.  This condition reduces to $\{1+2b^2(r_{h}^2- \sqrt{r_{h}^4+a^4}) \}
< 1$ that is satisfied since $ (r_{h}^2- \sqrt{r_{h}^4+a^4}) < 0$. The
condition is illustrated in Fig.\ref{biihfig4} where $\kappa_{(b)}(r_h)=
\frac{1}{2r_h} \{1+2b^2(r_{h}^2- \sqrt{r_{h}^4+a^4}) \} $.
 
(v) This prediction is about the behavior of $M^{(b)}_{hor}(r_h)$ and is
the only one that the EBI solution does not fulfil. Contrary to the EYM
case, the $M^{(b)}_{hor}(r_h)$, as a function of $r_h$ (fixed $b$), does
not increase monotonically, as can be seen in Fig. {\ref{biihfig2}} that
for $b=1.5$ and larger, $M^{(b)}_{hor}(r_h)$ has a minimum and then
increases monotonically. Asymptotically its slope tends to the one of the
bare black hole. We note that for the EYM solution this prediction was done
on the basis of two numerical cases ($n=1,2$) and for low $b$ the EBI
solution also satisfy this condition.

(vi) For any given $b$, the curve $M^{(b)}_{hor}(r_h)$ lies between the two
parallel lines $(r_h/2 -M^{(b)}_{sol})$ and $1/r_h$. Using the
corresponding expressions for $M^{(b)}_{hor}(r_h)$, Eq. (\ref{Mhor}) and
$M_{sol}^{(b)}=\frac{2q \sqrt{qb}}{3}{\rm K}(\frac{1}{2})$ and also Eq.
(\ref{relint}), this condition reduces to the fulfilment of the inequality

\begin{equation}
2q^2
\int^{\infty}_{r}{\frac{ds}{\sqrt{s^4+a^4}}}>b^2r^3[\sqrt{1+\frac{a^4}{r^4}}-1],
\label{ineq}
\end{equation}
that in fact is true as can be shown plotting both sides of the inequality.

\begin{figure}
\includegraphics[width=8.3cm,height=6cm]{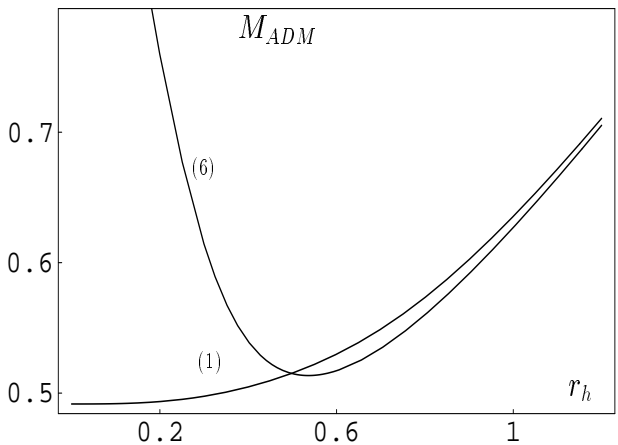}
\caption{\label{biihfig5}
It can be observed how the EBI ADM mass, $M_{ADM}^{(b)}(r_{h})$, show the
crossing of families corresponding to distinct BI parameter $b$. The plot
shows the cases $b=1$ and $b=6$ (in parenthesis). }
\end{figure}

Additionally the EBI ADM masses, $M_{ADM}^{(b)}(r_{h})$ show the crossing
of families corresponding to distinct BI parameter $b$, as can be seen in
the Fig. \ref{biihfig5}. This is also a feature in the EYM case.

\subsection{Final Remarks}
 
We have shown that the the static sector of the EBI theory is described by
the heuristic model for the colored black holes proposed by Ashtekar {\it
et al}. It is carried out provided that in the EBI theory there exist both
exact solutions: the colored black hole and the soliton like solution. In
such a manner that the predictions of the model can be demonstrated
analytically, and they are shown in plots as a better illustration. There
is only one prediction of the model that the EBI solution does not fulfil:
$M_{hor}^{(b)}(r_h)$ does not increase monotonically, except for low values
of $b$.

We also point out some differences in comparison with the EYM black hole
whose study lead to the model \cite{ulises}: the Abelian character of BI
theory vs. non-Abelian EYM and that BI theory is described with a continuos
parameter $b$ in contrast with the discreteness of the EYM parameter $n$.

In general it occurs that $M_{ADM}^{(b)} > M_{hor}^{(b)}$. Using the
corresponding expressions Eqs. (\ref{Madm}) and (\ref{Mhor}), this
inequality reduces to inequality (\ref{ineq}). Since the diference between
the hamiltonian horizon mass and the ADM mass can be seen as the energy
that is available for radiation to fall both into the black hole and to
infinity, then the nonzero value of the Hamiltonian could be an indication
of instability of the EBI solution. On this basis, one can conjecture that
the EBI black hole is unstable.

\begin{acknowledgments}
The author acknowledges discussions with U. Nucamendi.
\end{acknowledgments}


\begin{thebibliography}{99}

\leftmargin 2.5em
\bibitem{Born}
M. Born and L. Infeld, Proc. R. Soc. (London) {\bf A144}, 425 (1934).
\bibitem{GSP}
A. Garc\'{\i}a, H. Salazar and J. F. Pleba\~nski, Nuovo Cim. {\bf 84}
(1984), 65-90.
\bibitem{Geon}
M. Demianski, Found. of Phys. {\bf 16} (1986),187-190.
\bibitem{Ashtekar}
A. Ashtekar, S. Fairhurst and B. Krishnan, Phys. Rev. {\bf D 62}
(2000), 104025.
\bibitem{Sudarsky}
A. Ashtekar, A. Corichi and D. Sudarsky, Class.Quant.Grav. {\bf 18
}(2001), 919-940.
\bibitem{Breton}
N. Bret\'on, Class. and Quantum Grav. {\bf 19} (2002),601-612.
\bibitem{ulises}
A. Corichi, U. Nucamendi and D. Sudarsky, Phys. Rev. {\bf D 62} (2000),
044046.
\end{thebibliography}
\end{document}